\documentstyle[12pt,a4wide]{article}
\title{\mbox{\LARGE{On the Evolution of Primordial Magnetic Fields}}}
\author{
\mbox{\large{Konstantinos Dimopoulos}}
\thanks{e-mail: K.Dimopoulos@damtp.cam.ac.uk}
\vspace{0.2cm}\\
\mbox{\large{and}}
\vspace{0.2cm}\\
\mbox{\large{Anne--Christine Davis}}
\thanks{e-mail: A.C.Davis@damtp.cam.ac.uk}
\vspace{0.4cm}\\
\mbox{\normalsize{\em Department of Applied Mathematics and
Theoretical Physics,}}\\
\mbox{\normalsize{\em University of Cambridge, Silver Street,}}\\
\mbox{\normalsize{\em Cambridge, CB3 9EW, U.K.}}
}

\begin{document}
\begin{titlepage}
\maketitle
\begin{abstract}
A new mechanism for the evolution of primordial magnetic fields is
described and analysed. The field evolution is followed from the time
of its creation until the epoch of structure and galaxy formation. The
mechanism takes into account the turbulent behaviour of the early
universe plasma, whose properties determine strongly the evolution of
the field configuration. A number of other related issues such as the
case of an electroweak plasma are also considered. Finally, 
as an example, the mechanism is applied to specific models. 
\end{abstract}
\vspace{1cm}
\flushright{DAMTP-96-17}
\end{titlepage}

{\bf 1.}\ \ One of the most exciting astrophysical consequences of
phase transitions in the early universe is the possible creation of
primordial magnetic fields. The existence of a primordial magnetic
field could have significant effect on various astrophysical
processes. In particular it may be involved in the galaxy formation
process \cite{wasser}, \cite{peeb1} or in the generation of the
observed galactic magnetic fields. 

It is widely accepted that the galactic magnetic fields are generated
through a galactic dynamo mechanism. Here, a weak seed field is
exponentially amplified by the turbulent motion of ionized gas, which
follows the differential rotation of the galaxy \cite{zeld}, \cite{park}.
The currently observed magnetic field of the Milky Way and of nearby
galaxies is of the order of a $\mu Gauss$. If
the \mbox{e-folding} time is no more than the galactic rotation
period, \mbox{$\sim 10^{8}yrs $}, then, considering the galactic age,
\mbox{$\sim 10^{10}yrs $}, the seed field needed to produce a field of
the observed value is about \mbox{$\sim 10^{-19}Gauss $} \cite{zeld}
on a comoving scale of a protogalaxy (\mbox{$\sim 100\,kpc$}). Since
the gravitational collapse of the protogalaxies enchances their
frozen-in magnetic field by a factor of
\mbox{$(\rho_{G}/\rho_{0})^{2/3}\sim 10^{3}$}, where
\mbox{$\rho_{G}\sim 10^{-24}g\,cm^{-3}$} is the typical mass density
of a galaxy and \mbox{$\rho_{0}\simeq 2\times 10^{-29}\Omega
h^{2}g\,cm^{-3}$} is the current cosmic mass density, this seed
field corresponds to an rms field of the order of \mbox{$\sim
10^{-22}Gauss$} over the comoving scale of \mbox{$\sim 1\,M\!pc$}.
With the assumption that the rms field scales approximately 
as $a^{-2}$ with the expansion of the
universe, where \mbox{$a\propto t^{2/3}$} is the scale factor in the
matter era, we find that the required rms value of the seed field at
the time \mbox{$t_{eq}\sim 10^{11}sec$} of equal matter and radiation
densities is \mbox{$\sim 10^{-22}Gauss\times(t_{gc}/t_{eq})^{4/3}$}
\mbox{$\sim 10^{-20}Gauss$}, where \mbox{$t_{gc}\sim 10^{15}sec$} is
the time of the gravitational collapse of the galaxies \cite{wasser}.
Thus, an rms field of magnitude, \mbox{$B^{eq}_{rms}\sim
10^{-20}Gauss$} at $t_{eq}$ would be sufficient to seed the galactic
dynamo and generate the observed galactic magnetic fields. 

Various attempts have been made to produce a primordial field in the
early universe \cite{rest}. In most of the cases, though, the achieved
field appeared to be too weak or incoherent to seed the galactic
dynamo. In this letter, it is shown that the key issue, determining
the rms value of a primordial magnetic field at $t_{eq}$, is the
evolution of the correlated domains of the field, i.e. the growth of
the lengthscale over which the magnetic field is coherent. We develop
a detailed mechanism for the evolution of the magnetic field
configuration and we show that, in contrast to what is usually
assumed, the correlation length, in general, grows faster than the
scale factor $a$. This results in more coherent rms fields at the
epoch of galaxy formation.  

{\bf 2.}\ \ In the existing literature the entire magnetic field
configuration is taken to be comovingly frozen. As a result the rms
magnetic field is thought to evolve as $a^{-2}$, due to flux
conservation. 

The fact that the early universe is, with great accuracy, a perfect
conductor, ensures that magnetic flux is, indeed, conserved 
(in absense of dissipation) and,
therefore, the magnetic field can be thought to be ``frozen'' into the
plasma \cite{nuc1}, \cite{enqv}, over a certain scale. However,
the simplistic assumption that the correlated domains of the field
expand only due to the Hubble expansion does not take into account
that the faster expanding causal correlations, through electromagnetic
turbulence, could rearrange the field and correlate it on
comoving scales larger than its initial correlated domains. 
This is because, when two initially uncorrelated neighbouring domains
come into causal contact, the magnetic field around the interface is
expected to untangle and smooth, in order to avoid the creation of 
energetically unfavoured magnetic domain walls. In time the field
inside both domains ``aligns'' itself and becomes coherent over the
total volume. The velocity $v$, with which such a reorientation
occurs, is determined by the plasma, which carries the field and has
to reorientate its motion for that purpose.\footnote{Note that the
plasma does not have to be carried from one domain to another or get
somehow mixed. Also, conservation of flux is not violated with the
field's rearrangements, since {\em the field always remains frozen
into the plasma}, which is carried along.} 

Thus, the evolution of the correlation length is given by,

\begin{equation}
\frac{\mbox{d}\xi }{\mbox{d}t}=H\xi +v \label{dxi}
\end{equation}
where $\xi $ is the correlation length of the magnetic field
configuration, $H$ is the Hubble parameter and $v$ is the peculiar,
bulk velocity, determined, in principle, by the state of the plasma. 
 
From (\ref{dxi}) it is apparent that the correlated domains could grow
faster than Hubble expansion. Therefore, {\em the magnetic field
configuration is not necessarily comovingly frozen}. Indeed
we show that the domains can expand much faster than the universe,
resulting in large correlations of the field. 

Here it is important to point out our basic implicit assumption,
which concerns the damping of the small scale structure of the
magnetic field as the correlated domains expand. Indeed, although we
will not refer to specific damping mechanisms, {\em we assume that all
the Fourier modes of the magnetic field with wavelength smaller than
the dimensions of the correlated domains are fully damped} and, therefore,
the field is coherent inside these domains. This is equivalent to
assuming that the magnetohydrodynamic backreaction, which could
transfer power to the small scales at later times, is not effective.
This assumption enables us to avoid the intrinsic non-linearities of
the problem (by confining them into the damping mechanisms) and to
attempt a linear approach {\em without using perturbation theory}.
Of course, a more realistic picture would have to include some
transfer of power to the smaller than the correlation length scales,
especially the ones that are near the scale of the correlated domains.
In that sense our work can be viewed as {\em the optimum case}, i.e.
the limit of the fastest possible growth of the correlated domains. As
such we can still have predictive power by setting upper limits to the
strength and coherence of the magnetic field configuration at any
given time. 

In order to describe the evolution of the correlated domains one has
to determine the peculiar velocity $v$ of equation (\ref{dxi}). 
This primarily depends on the opacity of the plasma. 

If the plasma is opaque on the scale of a correlated domain, then
radiation cannot penetrate this scale and is blocked inside the plasma
volume. Consequently, the plasma is subject to the total magnetic
pressure of the magnetic field gradient energy. Therefore, this energy
dissipates through coherent magnetohydrodynamic oscillations, i.e.
Alfven waves. This is evident, since the coherent, bulk velocity of
the plasma motion is driven by the magnetic field so that \mbox{$\rho
v^{2}\sim B^{2}$}. Thus, there is equipartition of energy between the
coherent motion of the plasma and the magnetic filed\footnote{The
coherent plasma motion should not be confused with the thermal motion
where \mbox{$v_{t\!h}\simeq\sqrt{T/m}$}.}.
Consequently, in this case, the peculiar velocity of the magnetic field
reorientation, is the well known
Alfven velocity \cite {hogan},\footnote{Unless explicitly specified,
natural units are being used (\mbox{$\hbar =c=1$}).}

\begin{equation}
v_{A}\equiv
\frac{B_{cd}}{\sqrt{\rho }} \label{va}
\end{equation}
where $B_{cd}$ is the magnitude of the magnetic field inside a
correlated domain and $\rho $ is the {\em total} energy density of
the universe, since,
before $t_{eq}$, matter and radiation are strongly
coupled.\footnote{This coupling implies that any reorientation of the
momentum of matter has to drag radiation along with it. This
increases the inertia of the plasma, that balances the magnetic
pressure.}

Now, if the plasma is not opaque over the scale $\xi $ of a correlated
domain, then radiation can penetrate this scale and carry away
momentum, extracted from the plasma through Thomson scattering of the
photons. This subtraction of momentum is equivalent to an effective
drag force, \mbox{$F\sim\rho\,\sigma_{T}\,v_{T}\,n_{e}$} \cite{hogan}.
Balancing this force with the magnetic force determines the
``Thomson'' velocity over the scale $\xi $,

\begin{equation}
v_{T}\equiv\frac{v_{A}^{2} }{\xi n_{e}\sigma_{T} } \label{vt}
\end{equation}
where $v_{A}$ is the Alfven velocity, $n_{e}$ is the electron number
density and $\sigma_{T}$ is the Thomson cross-section.

Hence, for a non--opaque plasma the peculiar velocity
of the plasma reorientation is given by \cite{hogan},

\begin{equation}
v=\mbox{min}(v_{A},v_{T})
\label{v}
\end{equation}

In order to explore the behaviour of the opaqueness of the plasma, we
need to compare the mean free path of the photon \mbox{$l_{T}\sim
(n_{e}\sigma_{T})^{-1}$} to the scale $\xi $ of the correlated
domains. For realistic models, the correlated
domains remain opaque at least until the epoch \mbox{$t_{anh}\sim
0.1\,sec$} of electron--positron annihilation \mbox{($T\sim
1\,M\!eV$)}. The reason for this can be easily understood by
calculating $l_{T}$ before and after pair annihilation.

For \mbox{$T>1\,M\!eV$}, instead of the usual Thomson cross-section
$\sigma_{T}$, we  have the Klein-Nishina cross-section
\cite{KN},  

\begin{equation}
\sigma_{KN}\simeq \frac{3}{8}\,\sigma_{T}\,(\frac{m_{e}}{T})\,[\,\ln
\frac{2T}{m_{e}}+\frac{1}{2}\,]
\simeq 2.7\,(\frac{GeV}{T})\,\ln [\frac{T}{GeV}]\,GeV^{-2}
\label{sKN}
\end{equation}
where \mbox{$m_{e}\simeq 0.5\,GeV$}is the electron mass and
\mbox{$\sigma_{T}\simeq 6.65\times 10^{-25}cm^{2}\simeq
1707.8\,GeV^{-2}$}. The electron number density is given by \cite{KT},

\begin{equation}
n_{e}\simeq \frac{3}{4}\frac{\zeta(3)}{\pi^{2}}\,g_{e}T^{3}
\label{ne}
\end{equation}
where \mbox{$\zeta(3)\simeq 1.20206$} and
\mbox{$g_{e}=4$} are the internal degrees of freedom of
electrons and positrons.

From (\ref{sKN}) and (\ref{ne}) we find,

\begin{equation}
l_{T}\sim\frac{0.1\,GeV}{T^{2}}\;\;\;\;\;\;\;\;\;\;\mbox{for
$T>1\,M\!eV$}
\end{equation}
which at annihilation gives, \mbox{$l_{T}(t_{anh})\sim
10^{5}GeV^{-1}$}.

After annihilation the electron number density is given by
\cite{KT},

\begin{equation}
n_{e}\simeq 6\times 10^{-10}n_{\gamma }\simeq 1.44\times 10^{-10}T^{3}
\label{NE}
\end{equation}
where $n_{\gamma }$ is the photon number density given by,

\begin{equation}
n_{\gamma }\simeq \frac{\zeta(3)}{\pi^{2}}\,g_{\gamma }T^{3}
\label{ng}
\end{equation}
where \mbox{$g_{\gamma }=2$} are internal degrees of freedom of the
photon. 

With the usual value for $\sigma_{T}$ we obtain,
\begin{equation}
l_{T}\sim\frac{10^{6}GeV^{2}}{T^{3}}\;\;\;\;\;\;\;\;\;\;\mbox{for
$T<1\,M\!eV$}
\end{equation}

At annihilation the above gives, \mbox{$l_{T}(t_{anh})\sim
10^{15}GeV^{-1}$}. 

Hence, the mean free path of the photon at the time of pair
annihilation is enlarged by a factor of $10^{10}$! As a result,
$l_{T}$ is very likely to become larger than $\xi$ after $t_{anh}$.
If this is so, the Thomson dragging effect has to be taken into
account and the peculiar velocity of the plasma reorientation is given
by (\ref{v}).

In order to calculate the peculiar velocity it is necessary to compute
the Alfven velocity, which requires the knowledge of the magnetic
field value $B_{cd}$ inside a correlated domain. To estimate that we
assume that the magnetic flux, on scales very much larger than the
sizes of the correlated domains, is conserved, as implied by the
frozen--in condition.

Consider a closed curve $C$ in space, of lengthscale \mbox{$L>\xi$},
encircling an area $A$. Conservation of flux suggests
that the flux averaged mean magnetic field inside $A$ scales as
$a^{-2}$.  This implies that for the
field inside a correlated domain we have,
\mbox{$B_{cd}(L/\xi)^{-1}\propto a^{-2}$}. Since $C$ follows the
universe expansion \mbox{$L\propto a$}, with \mbox{$a\propto t^{1/2}$}.
Thus, for the radiation era, we obtain,

\begin{equation}
B_{cd}\,t^{1/2}\xi=K\Rightarrow B_{cd}=\frac{K}{t^{1/2}\xi}
\label{K}
\end{equation}
where $K$ is a constant to be evaluated at any convenient time.
We will show that the correlation length grows at least as fast as the
universe expands. This implies that the magnetic field inside a
correlated domain dilutes at least as rapidly as $a^{-2}$ for the
radiation era.

Subsituting the above into (\ref{va}) we find,

\begin{equation}
v_{A}\sim 10\,\frac{K}{m_{pl}}\,\frac{t^{1/2}}{\xi}
\label{vA}
\end{equation}
where we have also used the well known relation,

\begin{equation}
t\simeq 0.3\,g_{*}^{-1/2}(\frac{m_{pl}}{T^{2}})
\label{tT}
\end{equation}
where \mbox{$m_{pl}\simeq 1.22\times 10^{19}GeV$} is the Planck mass
and $g_{*}$ is the number of particle degrees of freedom, which, for
\mbox{$T<1\,M\!eV$} is 3.36 \cite{KT}.
\footnote{In natural units
\mbox{$G=m_{pl}^{-2}$}.} 

Solving the evolution equation (\ref{dxi}) with \mbox{$a\propto
t^{1/2}$} in the case that \mbox{$v=v_{A}$} gives,

\begin{equation}
\xi(t)^{2}=(\frac{t}{t_{i}})\,\xi_{i}^{2}+4v_{A}(t)\,\xi(t)\,
t\,(1-\sqrt{\frac{t_{i}}{t}}\,) 
\label{xivA}
\end{equation}
where $\xi_{i}$ is the correlation length of the field at the time
$t_{i}$. The first term of (\ref{xivA}) is due to the Hubble
expansion, whereas the second term is due to the peculiar velocity. 

In the case of \mbox{$v=v_{T}$}, for
\mbox{$t>t_{anh}$}, using (\ref{NE}), (\ref{ng}) and the usual value
of $\sigma_{T}$, (\ref{vt}) gives,

\begin{equation}
v_{T}=D\,\frac{t^{5/2}}{\xi^{3}}
\label{vT}
\end{equation}

where

\begin{equation}
D\sim 10^{-57}K^{2}GeV^{-3/2}
\label{D}
\end{equation}

Using (\ref{vT}), the evolution equation (\ref{dxi}) gives,

\begin{equation}
\xi(t)^{4} = (\frac{t}{t_{i}})^{2}\,\xi_{i}^{4}+
\frac{8}{3}\,v_{T}(t)\,\xi^{3}t\,[1-(\frac{t_{i}}{t})^{3/2}]
\label{xivT}
\end{equation}

The evolution of the correlation length of the magnetic field
configuration is described initially by the Alfven expansion equation
(\ref{xivA}) until the moment when \mbox{$\xi\sim l_{T}$}. From then
on the growth of $\xi$ continues according either to (\ref{xivA}) or
to (\ref{xivT}), depending on the relative magnitudes of the
velocities $v_{A}$ and $v_{T}$. Using the above, the scale
$\xi_{eq}$ of the correlated domains at $t_{eq}$ can be estimated.
With a suitable averaging procedure, this this can be used to
calculate the rms magnetic field  over the protogalactic comoving
scale at the time when structure formation begins. 

{\bf 3.}\ \ An important issue, which should be considered, is the
diffusion length of the freezing of the field. Indeed, the assumption
that the field is frozen into the plasma corresponds to neglecting the
diffusion term of the magnetohydrodynamical induction equation
\cite{jack}, 

\begin{equation}
\frac{\partial {\bf B}}{\partial t}=\nabla\times({\bf v}\times{\bf B})
+ \sigma^{-1}\nabla^{2}{\bf B} \label{XB}
\end{equation}
where {\bf v} is the plasma velocity and $\sigma$ is the
conductivity. In the limit of infinite conductivity 
the diffusion term of (\ref{XB}) vanishes and the
field is frozen into the plasma on all scales. However, if
$\sigma$ is finite then spatial variations of the magnetic field
of lengthscale $l$ will decay in a diffusion time,
\mbox{$\tau\simeq\sigma l^{2}$} \cite{jack}. Thus, the field at a
given time $t$ can be considered frozen into the plasma only over the
diffusion scale, 

\begin{equation}
l_{d}\sim\sqrt{\frac{t}{\sigma}} \label{ld}
\end{equation}

If \mbox{$l_{d}>\xi$}, the magnetic field
configuration is expected, in less than a Hubble time, to become
smooth on scales smaller than $l_{d}$. Thus, it is more
realistic to consider a field configuration with
coherence length $l_{d}(t)$ and magnitude of the coherent magnetic
field $\overline{B_{cd}}$, where \mbox{$\overline{B_{cd}}=B_{cd}/N$}
is the flux-averaged initial magnetic field over \mbox{$N\equiv
l_{d}/\xi$} number of domains.

An estimate of the plasma conductivity is necessary
to determine the diffusion length. The current density in the
plasma is given by, \mbox{${\bf J}=ne{\bf v}$}, where $n$ is the
number density of the charged particles. The velocity {\bf v} acquired
by the particles due to the electic field {\bf E}, can be estimated as
\mbox{${\bf v}\simeq e{\bf E}\tau_{c}/m$}, where $m$ is the particle
mass and \mbox{$\tau_{c}=l_{m\!f\!p}/v$} is the timescale of collisions.
Since the mean free path of the particles is given by,
\mbox{$l_{m\!f\!p}\simeq 1/n\sigma_{c}$}, the current density is, 
\mbox{${\bf J}\simeq e^{2}{\bf E}/mv\sigma_{c}$}, where $\sigma_{c}$
is the collision cross-section of the plasma particles. Comparing with
Ohm's law gives for the conductivity \cite{jack}, \cite{plasma},

\begin{equation}
\sigma\simeq\frac{e^{2}}{mv\sigma_{c}}=
\frac{\omega_{p}^{2}}{4\pi\nu_{c}}
\label{sigma}
\end{equation}
where \mbox{$\omega_{p}\equiv (\frac{4\pi ne^{2}}{m})^{1/2}$} is the
plasma frequency and \mbox{$\nu_{c}=nv\sigma_{c}$} is the frequency of
collisions. The collision cross-section is given by the Coulomb formula
\cite{plasma},

\begin{equation}
\sigma_{c}\simeq\frac{e^{4}}{T^{2}}\,\ln\Lambda
\label{coulob}
\end{equation}
where \mbox{$\ln\Lambda\simeq\ln(e^{-3}\sqrt{T^{3}/n})$} is the Coulomb
logarithm. Thus, the behaviour of the
conductivity depends crucially on the temperature.

For low temperatures, \mbox{$T<m_{e}\simeq
1\,M\!eV$} (i.e. after $t_{anh}$), 
the velocity of the electrons is, \mbox{$v\sim\sqrt{T/m_{e}}$}. Thus, from
(\ref{sigma}) and (\ref{coulob}) the conductivity is given by,

\begin{equation}
\sigma\sim\frac{1}{e^{2}}\sqrt{\frac{T^{3}}{m_{e}}}\,\frac{1}{\ln\Lambda}
\label{sL}
\end{equation}

For high temperatures, \mbox{$T\gg m$},\vline\ 
\footnote{At not extremely high
temperatures the roles of the electrons and the protons may be
reversed \cite{hary} which would imply that \mbox{$m=m_{p}$} ($m_{p}$
being the proton mass) instead of $m_{e}$. This would decrease the
conductivity by a factor of \mbox{$(m_{e}/m_{p})^{1/2}\simeq 0.02$}. At
even higher temperatures \mbox{$T\gg m_{p}$}, thermal corrections become
dominant.} 
(\ref{ne}) suggests that, \mbox{$\ln\Lambda\sim
1$}. Also, the mass of the plasma particles is dominated by
thermal corrections, i.e. \mbox{$m\sim T$}, and \mbox{$v\sim 1$}.
Consequently, in this case, (\ref{sigma}) and (\ref{coulob}) give for the
conductivity,
\begin{equation}
\sigma\sim c\,\frac{T}{e^{2}}
\label{s}
\end{equation}
where $c$ is a numerical factor of order unity.\footnote{As it is
shown in \cite{ahonen} the main contribution to the early universe
conductivity is from leptonic interactions. $c$ is found to be a
slowly increasing function of temperature (since at low temperatures
most of the leptonic species heve underwent a pair--annihilation
period). It is shown that $c$ ranges from 0.07 at \mbox{$T\sim
100\,M\!e\!V$} to 0.6 at \mbox{$T\sim 100\,G\!e\!V$}. For even higher
temperatures $c$ approaches the textbook estimate \mbox{$c\simeq 1.3$}
\cite{text} for relativistic electron scattering off heavy ions.}

Using the above results we can estimate the diffusion length. Indeed,
from (\ref{ld}), (\ref{sL}) and (\ref{s}) we obtain,

\begin{equation}
l_{d}\sim\left\{ \begin{array}{lr}
10^{8}GeV^{1/2}T^{-3/2} & \;\;\;\;\;T\geq 1\,M\!eV \\
 & \\
10^{8}GeV^{3/4}T^{-7/4} & \;\;\;\;\;T<1\,M\!eV
\end{array}\right.
\label{ldT}
\end{equation}

An important point to stress is that the diffusion
length is also increasing with time. If \mbox{$l_{d}>\xi$} then the
size of the correlated domains is actually determined by the diffusion
length and it is the growth of the later that drives the evolution of
the magnetic field configuration. 

Another lengthscale necessary to consider is the magnetic Jeans
length, given by \cite{peeb1} (see also \cite{nuc2} and
\cite{olinto}). This lengthscale is a measure of the dissipation of
the field after $t_{eq}$. Indeed, in order for the field to have any
astrophysical implications in structure formation, it needs to be
coherent over, 

\begin{equation}
\lambda_{_{B}}^{eq}\sim\frac{B^{eq}}{2\rho}
\,m_{pl}
\label{lJ}
\end{equation}
(i.e. \mbox{$\xi_{eq}\geq\lambda_{_{B}}^{eq}$}). If this is
not the case then field oscillations dissipate the energy of the field
and lead to its decay. If the field {\em is} coherent over the above
lengthscale then it can have an accumulative effect and result in
density instabilities, which can, by themselves, lead to structure and
galaxy formation \cite{wasser}, \cite{olinto}.

{\bf 4.}\ \ Following the above analysis, the strength and coherence
of the magnetic field configuration at any stage of its
evolution\footnote{though not after $t_{eq}$}, can
be calculated if the intitial values of the field and the correlation
length are given. 

Being interested in the galactic magnetic fields, we would attempt to
calculate the rms field on the scale of a protogalaxy \mbox{$\sim
1\,M\!pc$} at the time $t_{eq}$, when structure formation begins.
To do so we need to employ a suitable averaging procedure.

Choosing the magnetic field as a stochastic variable, Enqvist and
Olesen \cite{enqv} have shown that the root mean square value of the
field would behave as,

\begin{equation}
B_{rms}\equiv \sqrt{\langle B^{2}\rangle}=\frac{1}{\sqrt{N}}\,B_{cd}
\label{brms}
\end{equation}
where $N$ is the number of correlation lengthscales over which the
field is averaged. At $t_{eq}$ we have,
\begin{equation}
N=\frac{L_{g}^{eq}}{\xi_{eq}}
\label{N}
\end{equation}
where $L_{g}^{eq}$ is the protogalactic scale at $t_{eq}$, for which
we find, \mbox{$L_{g}^{eq}\sim (t_{eq}/t_{pr})^{2/3}L_{g}^{pr}\sim
10\,pc$}, where \mbox{$t_{pr}\sim 10^{18}sec$} is the present time and
\mbox{$L_{g}^{pr}=1\,M\!pc$}. 

At this point it should be mentioned that, in the above treatment, the
rms value of the field has been computed as a line average. The
averaging procedure could have an important effect on the results and
has to be considered carefully. One argument in favour of
line-averaging is that the observed galactic magnetic fields have been
measured using the Faraday rotation of light spectra, which is also a
line (line of sight) computation. If we assume that the ratio of the
galactic dynamo seed field and the currently observed galactic field is
independent of the averaging procedure then this would suggest
line-averaging is required for the computation of the primordial rms
field. However, the nonlinearity of the dynamo process as well as the
rather poor knowledge we have for galaxy formation make such an
assumption non-trivial. In any case, apart from the above, there seem
to be no other argument in favour of a particular averaging procedure.
Therefore, using line-averaging could be the safest choice.

Here it is also important to point out that 
{\em line averaging just gives an estimate of the rms field and does
not correspond to any physical process}. Thus, it should not be
confused with flux averaging which {\em does} correspond to a physical
process, that of the field untangling, and is so in order to preserve
flux conservation on scales larger than the diffusion length.

{\bf 5.}\ \ If the magnetic field is produced at very early times e.g.
during an inflationary era, electroweak unification needs to be taken
into account. If the original field is created before the electroweak
transition then, assuming that it becomes ``frozen'' into the
electroweak plasma is non--trivial. Indeed, during
the electroweak era, since the electroweak symmetry group 
\mbox{$SU(2)\times U(1)_{Y}$} is unbroken, there are four
apparent ``magnetic'' fields, three of which are non-Abelian.
It would be more precise, then,  to refer only to
the Abelian (hypercharge) part of the magnetic field, which satisfies
the same magnetohydrodynamical equations as the Maxwell field of
electromagnetism. The non-Abelian part of the field may not influence
the motion of the plasma due to the existence of a temperature
dependent magnetic mass, \mbox{$m_{B}\approx 0.28g^{2}T$} \cite{mB}, 
which could screen the field. Then, the motion of the plasma is 
determined primarily by the Abelian field and can reach a
selfconsistent pattern, which will ``lock'' onto the field in the same
way as in electromagnetism. 

The condition for this screening to be effective is obtained by
comparing the screening length \mbox{$r_{S}\sim m_{B}^{-1}$} of the
non-Abelian magnetic fields with the Larmor radius of the plasma
motion \mbox{$r_{L}\sim\frac{mv}{gB}$}, where
\mbox{$m\sim\sqrt{\alpha}\,T$} (\mbox{$\alpha =g^{2}/4\pi$}) is the
temperature induced physical mass of the plasma particles, $g$ is the
gauge coupling (charge) and $v$ is the plasma particle velocity. 
Assuming thermal velocity distribution, i.e. \mbox{$mv^{2}\sim T$},
suggests,

\begin{equation}
R\equiv\frac{r_{L}}{r_{S}}\sim 10^{-2}\frac{T^{2}}{B_{cd}}
\end{equation}

If \mbox{$R\geq 1$} then the restriction
to the Abelian (Hypercharge) part of the magnetic field is justified. 
This restriction would not cause any significant change
to the final magnitude of the ``electromagnetic'' magnetic field,
since, at the electroweak transition, the hypercharge symmetry
projects onto the photon through the Weinberg angle,
\mbox{$\cos\theta_{W}\approx 0.88$}. If, however, \mbox{$R<1$} then
the non-Abelian fields do affect the plasma motion and should be taken
into account.
Since, \mbox{$T\propto a^{-1}$} and $B_{cd}$ falls at least as rapid
as $a^{-2}$, $R$ is in general an increasing function of time. Thus,
the constraint has to be evaluated at the time of creation of the
magnetic field configuration. 

{\bf 6.}\ \ In determining the rms value of the field at $t_{eq}$ one
has also to take into account a number of constraints regarding its
strength. An obvious requirement is that the magnetic field should not
dominate the energy density of the universe. The expansion of the
universe dilutes the energy density of the magnetic field,
\mbox{$\rho_{B}=B_{cd}^{2}/8\pi$},  
inside a correlated domain more effectively than the radiation
density, which scales as $a^{-4}$.
Therefore, it is sufficient to ensure that
$\rho_{B}(t)$ is less than the energy density $\rho(t)$ of radiation
at the time $t_{i}$ of the formation of the magnetic field configuration. 
Thus, the constraint reads,

\begin{equation}
\rho_{B}(t_{i})\leq\rho(t_{i})\Rightarrow
B_{cd}^{i}\leq\frac{\sqrt{3}}{2}\,\frac{m_{pl}}{t_{i}}
\end{equation}

Another constraint comes from nucleosynthesis. This has been
studied in detail by Cheng {\em et al.} \cite{nuc1}. They conclude
that, at \mbox{$t_{nuc}\sim 1\,sec$}, the magnetic field should not be
stronger than,

\begin{equation}
B^{nuc}\leq 10^{11}Gauss
\end{equation}
on a scale larger than \mbox{$\sim 10^{4}cm$}. A more recent
treatment by Kernan {\em et al.} \cite{nuc2} relaxes the bound by about
an order of magnitude, \mbox{$B^{nuc}\leq
e^{-1}(T_{\nu}^{nuc})^{2}\sim 10^{12}Gauss$}, where $T_{\nu}$ is the
neutrino temperature and $e$ is the electric charge. This bound is
valid over all scales. Similar results are also reached by Grasso and
Rubinstein \cite{grasso}. 

Finally, an additional consideration would be the lower bound due to
the galactic dynamo requirements, which, as already explained, demands
a field of magnitude,

\begin{equation}
B^{eq}\geq 10^{-20}Gauss
\end{equation}
over the comoving scale of \mbox{$1\,M\!pc$}. 

Our field evolution mechanism results in a more coherent magnetic
field at $t_{eq}$ than previously considered.
As a result the rms field produced by the mechanisms in \cite{rest}
could also be of greater strength. In some cases, the
achieved rms magnetic field could be strong enough to dispense with the
galactic dynamo. 

At this point we should mention that there is some scepticism
regarding the galactic dynamo \cite{sofue}, mostly due to the fact
that, until now, there is no consistent dynamo model \cite{wein}.
If there is no dynamo
mechanism, the growth of the galactic magnetic field is only due to
winding and, therefore, is linear (in contrast to exponential). 
Some authors believe that this linear amplification of the field would
be enough to account for the currently observed galactic magnetic
field, considering also the additional enchancement by line dragging
during the gravitational collapse. It has been argued 
that, if this is the case, a seed field of the order of,
\mbox{$B^{eq}\sim 10^{-9}Gauss$} would be sufficient.
The gravitational collapse would also sweep
the intergalactic field lines into the galaxies leaving a relatively
weak intergalactic field, in agreement with observations, which find
that the intergalactic field is less than \mbox{$10^{-9}Gauss$}.

{\bf 7.}\ \ To demonstrate the efficiency of the mechanism, in terms
of magnitude and coherency, we will apply it to two
toy-models of primordial magnetic field generation. For each one of
these models the initial correlation length and the time of formation
of the initial primordial magnetic field configuration is specified
in such a way that it could correspond to a realistic situation. The
initial magnitude of the generated magnetic field, though, is treated
as a free parameter. In both cases we generate the field at a phase
transition, when the universe is out of thermal equilibrium and,
thus, the creation of such a field is acceptable \cite{critr}.

\bigskip

CASE 1. {\em At the electroweak phase transition.}

\bigskip

In general it is thought that the electroweak transition
is weakly first order, that is it occurs through bubble nucleation.
A natural choice for the correlation length of a magnetic field, that
is created at the transition, would be the bubbles' size at the time
of their collision. 
Using typical values for the parameters, we find,
\mbox{$\xi_{i}\sim (v_{w}/\Gamma)^{1/4}\sim 10^{10}GeV^{-1}$}
\mbox{$\sim 10^{-3}H_{ew}^{-1}$} \cite{martin}, where 
\mbox{$v_{w}\sim 0.1$} is the velocity of the bubble walls,
\mbox{$\Gamma\sim 10^{-41}GeV^{4}$} is the bubble nucleation rate
per unit time per unit volume \cite{turok} and \mbox{$H_{ew}^{-1}\sim
t_{ew}\sim 10^{13}GeV^{-1}$} is the horizon size at the time of the
transition. We assume that, by some mechanism, an initial magnetic
field, \mbox{$B_{cd}^{i}\sim 10^{Z}Gauss$} is generated at the
transition, where $Z$ is a free parameter. Using the above initial
conditions we can explore the evolution of such a field.

From (\ref{K}) we find that,

\begin{equation}
K\sim 10^{Z-4}GeV^{1/2}
\end{equation}

Inserting the above into (\ref{vA}) we find,

\begin{equation}
v_{A}\sim 10^{Z-22}GeV^{-1/2}\,\frac{t^{1/2}}{\xi}
\end{equation}

Using this in (\ref{xivA}) we can estimate the correlation length
$\xi_{anh}$ at the time $t_{anh}$ of pair annihilation,

\begin{equation}
\xi_{anh}\sim\left\{ 
\begin{array}{lr}
10^{15}GeV^{-1} & \;\;\;\;\;Z\leq 17\\
 & \\
10^{Z/2+7}GeV^{-1} & \;\;\;\;\;Z>17
\end{array}\right.
\end{equation}

The above suggests that the Alfven expansion
dominates only for \mbox{$Z>17$}. Otherwise it is the Hubble term that
determines the evolution of the correlation length. Using (\ref{ldT})
we can compute the diffusion length at $t_{anh}$. We find that
\mbox{$\xi_{anh}\gg \l_{d}^{anh}\sim 10^{13}GeV^{-1}$}, i.e. the
correlation length is always larger than the diffusion length at that
time. 

From (\ref{vA}) and (\ref{vT}) it can be easily verified that, for all
$Z$, \mbox{$v_{A}(t_{anh})>v_{T}(t_{anh})$}. Thus, from the time of
pair annihilation the correlation length evolves according to
(\ref{xivT}). Following its evolution likewise it can be shown that
the Thomson expansion cannot compete with the Hubble one. Thus, the
evolution of the correlation length is driven by the expansion of the
universe. At $t_{eq}$ we find,

\begin{equation}
\xi_{eq}\sim\left\{
\begin{array}{lr}
10^{21}GeV^{-1} & \;\;\;\;\;Z\leq 17\\
 & \\
10^{Z/2+13}GeV^{-1} & \;\;\;\;\;Z>17
\end{array}\right.
\end{equation}

The diffusion length at $t_{eq}$ is found by (\ref{ldT}) to be,
\mbox{$l_{d}^{eq}\sim 10^{23}GeV^{-1}$}. Thus, the above correlation
length is larger than the diffusion length only if \mbox{$Z>20$}. 
Therefore, the actual size of the correlated domains at $t_{eq}$
is given by,

\begin{equation}
\xi_{eq}\sim\left\{
\begin{array}{lr}
l_{d}^{eq}\sim 10^{23}GeV^{-1} & \;\;\;\;\;Z\leq 20\\
 & \\
10^{Z/2+13}GeV^{-1} & \;\;\;\;\;Z>20
\end{array}\right.
\end{equation}

With the use of the above and (\ref{K}) we find,

\begin{equation}
B_{cd}^{eq}\sim\left\{
\begin{array}{lr}
10^{Z-25}Gauss & \;\;\;\;\;Z\leq 20\\
 & \\
10^{Z/2-15}Gauss & \;\;\;\;\;Z>20
\end{array}\right.
\end{equation}

Also, from (\ref{N}) we have,

\begin{equation}
N\sim\left\{
\begin{array}{lr}
10^{10} & \;\;\;\;\;Z\leq 20\\
 & \\
10^{20-Z/2} & \;\;\;\;\;Z>20
\end{array}\right.
\end{equation}

Using the above (\ref{brms}) gives,

\begin{equation}
B_{rms}^{eq}\sim\left\{
\begin{array}{lr}
10^{Z-30}Gauss & \;\;\;\;\;Z\leq 20\\
 & \\
10^{3Z/4-25}Gauss & \;\;\;\;\;Z>20
\end{array}\right.
\end{equation}

By enforcing the various constraints we can specify the acceptable
range of the final rms magnetic field strength and the corresponding
necessary initial field.

The requirements of the galactic dynamo suggest that \mbox{$Z\geq
10$}. The energy density constraint demands that \mbox{$Z\leq 26$} and
the nucleosynthesis constraint requires \mbox{$Z\leq 28$}. Thus, the
acceptable range of values for the magnetic field is,

\begin{eqnarray}
10\;\;\; & \leq Z\leq & \;\;\;26\nonumber\\
 & & \label{Bew}\\
10^{-20}Gauss & \leq B_{rms}^{eq}\leq & 10^{-6}Gauss\nonumber
\end{eqnarray}

Checking with the magnetic Jeans length we also find that the magnetic
field could influence the structure formation process only if
\mbox{$Z<13$}. Finally, from (\ref{Bew}) it is apparent that for some
parameter space the galactic dynamo is not even required since an rms
field of strength up to a $\mu Gauss$ at $t_{eq}$ can be achieved. In
fact, we can dispense with the galactic dynamo for \mbox{$Z>21$}.

\bigskip

CASE 2: {\em At grand unification.}

\bigskip

According to most scenaria for the breaking of grand unification, this
occurs at a temperature \mbox{$T_{GUT}\sim 10^{16}GeV$}. Since not
many details are specified for the nature of this phase transition the
safest choice of a typical lengthscale is the horizon itself. Thus,
for the initial correlation length we choose, \mbox{$\xi_{i}\sim
t_{i}\sim 10^{-15}GeV^{-1}$}. Similarly with the electroweak
treatment, we assume that, at the GUT transition\footnote{GUT here 
stands for Grand Unified Theory.} a magnetic field of
strength, \mbox{$B_{cd}^{i}\sim 10^{W}Gauss$} is generated, where $W$
is treated as a free parameter.

In the same way as in the electroweak case we have,

\begin{equation}
K=10^{W-43}GeV^{1/2}
\end{equation}
with the use of which the correlation length at pair annihilation is, 

\begin{equation}
\xi_{anh}\sim\left\{
\begin{array}{lr}
l_{d}^{anh}\sim 10^{13}GeV^{-1} & \;\;\;\;\;W<52\\
 & \\
10^{W/2-13}GeV^{-1} & \;\;\;\;\;W\geq 52
\end{array}\right.
\end{equation}

Again, after pair annihilation it can be shown that
\mbox{$v_{A}(t_{anh})>v_{T}(t_{anh})$}. Thus the evolution of the
correlation length continues with Thomson expansion. This time,
however, the Thomson velocity is high enough to dominate the Hubble
expansion rate. The resulting corrlation lengthscale at $t_{eq}$ is,

\begin{equation}
\xi_{eq}\sim\left\{
\begin{array}{lr}
l_{d}^{eq}\sim 10^{23}GeV^{-1} & \;\;\;\;\;W<54\\
 & \\
10^{W/2-4}GeV^{-1} & \;\;\;\;\;W\geq 54
\end{array}\right.
\end{equation}

Using the above we can calculate $B_{cd}^{eq}$ and $N$ in a similar
way as in the electroweak case. Then it can immediately be shown that,

\begin{equation}
B_{rms}^{eq}\sim\left\{
\begin{array}{lr}
10^{W-69}Gauss & \;\;\;\;\;W<54\\
 & \\
10^{3W/4-55}Gauss & \;\;\;\;\;W\geq 54
\end{array}\right.
\end{equation}

By employing the various constraints on the above result we can
determine the acceptable range of the magnetic field rms values.
The galactic dynamo requirements suggest, \mbox{$W\geq 49$}, the
energy density constraint demands, \mbox{$W\leq 54$}, the
nucleosynthesis constraint requires, \mbox{$W\leq 66$} and the
non-Abelian constraint sets, \mbox{$W\leq 50$}. Thus, the acceptable
range is very small, \mbox{$49\leq W\leq 50$} and corresponds to an
rms magnetic field of order, \mbox{$B_{rms}^{eq}\sim
10^{-(19-20)}Gauss$}, which could just about seed the galactic dynamo
mechanism. Of course, one can imagnine that, if the non-Abelian
constraint is violated this should not necessarily mean that there
will not be any surviving magnetic field. For instance, it is highly
probable that, if the non-Abelian fields do influence the plasma
motion they could perturb it in such a way that the magnetic field
strength would diminish enough for the Larmor radious to increase over
the non-Abelian screening length (with some of the magnetic energy being
thermalized into the plasma). Thus, in this way stronger magnetic
fields could survive the non-Abelian era. Moreover, such a mechanism
could ``cut-down'' the initial strength of a primordial field and
reduce it enough never to challenge the nucleosynthesis and energy
density constraints. In that sense the non-Abelian stage of evolution
of the field could expand the upper bound for $W$ instead of reducing it
to as low as 50. Still, it is difficult to imagnine a way of producing
a stronger than $10^{-19}Gauss$ field, and in that sense
the GUT-transition provides a narrow window towards a galactic seed
field. By checking with the magnetic Jeans length, though, it can be
shown that such a field would also influence the structure formation
process and, therefore, it could have additional astrophysical effects.

The above was just a toy--model analysis of our mechanism. The
mechanism can be applied to a variety of models. A more detailed and
complicated example of such an application is given in \cite{mine},
where a primordial magnetic field  is created at the breaking of grand
unification during inflation, in such a way that the narrow window of
the GUT-case is achieved in a natural and realistic way.

{\bf 8.}\ \ In conclusion, we have analysed the evolution of
primordial magnetic fields and shown that, when the effects of the
surrounding plasma are taken into account, the correlation length of
the field configuration grows faster than the Hubble expansion. This
results into a more coherent magnetic field than
previously thought. There is some similarity with the recent work of
Brandenburg {\em et. al.} \cite{theirs}, who describe the plasma with
relativistic magnetohydrodynamic equations and turbulent behaviour. 
However, they are unable to solve their equations analytically, and,
instead, they attempt a numerical analysis in 2+1 dimensions. 
This also results in a faster growth of the coherence length.
Still, they do not take into account all the effects of the
plasma on the magnetic field configuration, such as, for example, 
the Thomson scattering effect.

We would like to thank K. Bajers, B. Carter, N. Turok, T. Vachaspati
and especially M.J. Rees for dicussions. 
This work was partly supported by PPARC, the Greek
State Scholarships Foundation (I.K.Y.) and
the E.U. under the HCM program (CHRX-CT94-0423).

\end{document}